\documentclass[preprint,prb,aps,showpacs]{revtex4}
\begin{document}
\title{Modified Zakharov equations for plasmas with a quantum correction}
\author{L.~G.~Garcia,
F.~Haas, L.~P.~L.~de Oliveira and J.~Goedert\footnote{\{\tt
garcia, ferhaas, luna, goedert\}@exatas.unisinos.br}}
\affiliation{Universidade do Vale do Rio dos Sinos -- UNISINOS \\
Aa. Unisinos, 950\\
93022-000 S\~ao Leopoldo, RS, Brazil}
\begin{abstract}
Quantum Zakharov equations are obtained to describe the
nonlinear interaction between quantum Langmuir waves and quantum
ion-acoustic waves. These quantum Zakharov equations are
applied to two model cases, namely the four-wave interaction and the
decay instability. In the case of the four-wave instability,
sufficiently large quantum effects tend to suppress the
instability. For the decay instability, the quantum Zakharov
equations lead to results similar to those of the classical
decay instability except for quantum correction terms in the
dispersion relations. Some considerations regarding the nonlinear
aspects of the quantum Zakharov equations are also offered.
\end{abstract}


\maketitle

\newpage

\section{Introduction}
The importance of quantum effects in ultra-small electronic
devices,\cite{markowich} in dense astrophysical plasma systems
\cite{jung} and in laser plasmas\cite{bonitz} have produced an
increasing interest on the investigation of the quantum
counterpart of some of the classical plasma physics phenomena. For
instance, quantum plasma echoes,\cite{mf} the expansion of a
quantum electron gas into vacuum,\cite{mola} the quantum two and
three stream instabilities\cite{hmf} and the quantum Landau
damping \cite{suh}
%
%
have been the subject of recent
investigations. Also, quantum methods like the Wigner-Moyal
transform have been used in the treatment of the Landau damping of
classical partially incoherent Langmuir waves.\cite{fedele}
Quantum models like the Schr\"odinger-Poisson system have also
been used, through the correspondence principle, for the numerical
simulation of the Vlasov-Poisson system.\cite{bertrand}

In this context, a mathematical formulation, based on the quantum
hydrodynamical model for charged particle systems,\cite{gardner,
man} was introduced to study the quantum version of the low
frequency ion-acoustic waves.\cite{haas2} In micro-electronics,
the quantum hydrodynamical model describes\cite{10} negative
differential resistance associated to resonant tunnelling diodes.
It can also model ultra-small high--electron--mobility
transistors.\cite{20} The quantum hydrodynamical model for charged
particle systems was also successfully used for the description of
quantum dissipation,\cite{lopez} under the same closure hypothesis
as that adopted by Manfredi and Haas.\cite{man} In the case of
quantum ion-acoustic waves,\cite{haas2} several features of pure
quantum origin were observed for the linear, weakly nonlinear and
fully nonlinear waves. The linear quantum ion-acoustic waves are
described by a dispersion relation which tends to the classical
dispersion relation as quantum effects goes to zero, in accordance
with the correspondence principle. The weakly nonlinear quantum
ion-acoustic waves are described by a modified Korteweg-de Vries
equation depending on a scaled $\hbar$ parameter. Finally, the
fully nonlinear quantum ion-acoustic waves can have a coherent,
periodic pattern, not present in the classical case. This points
to the intrinsically more reversible features of quantum plasmas,
as seen for instance in quantum echoes\cite{mf} and coherent
patterns in the quantum two stream instabilities\cite{hmf}.
%
%

The purpose of the present paper is to continue this investigation
by studying the nonlinear coupling between the quantum
ion-acoustic waves and the quantum Langmuir waves. At the
classical level, a set of coupled nonlinear wave equations
describing the interaction between high frequency Langmuir
waves and low frequency ion-acoustic waves was first derived by
Zakharov.\cite{zak} Since then, this system have been the subject
of a large number of studies.\cite{thorn} In one-dimension, the
Zakharov equations can be written (in normalized units) as
\begin{eqnarray}
\label{eq00}
&&i{\partial E\over\partial t}+{\partial^2 E \over\partial
x^2}=nE\,,\\
\label{eq01}
&&{\partial ^2n\over\partial t^2}-{\partial^2n\over\partial
x^2}=
{\partial^2|E|^2\over\partial x^2}\,,
\end{eqnarray}
where $E$ is the envelope of the high frequency electric field and
$n$ is the plasma density measured from its equilibrium value. The
system (\ref{eq00}-\ref{eq01}) can be derived from a hydrodynamic
description of the plasma\cite{thorn,nichol} by distinguishing two
different time scales, the slow time scale of the ions and the
fast time scale of the electrons. The low mobility of the ions as
compared to that of the electrons justifies this kind of
treatment. Since the Landau damping of the Langmuir waves is
neglected in the fluid description, the model
(\ref{eq00}-\ref{eq01}) is restricted by the condition $k\ll k_D$,
where $k$ is the wavenumber and $k_D$ is the Debye wavenumber.
Also, a weak turbulence condition is to be satisfied.\cite{thorn}

In this paper, modified Zakharov equations are obtained by use of
a quantum fluid approach. Specifically, we assume a two species,
one-dimensional quantum plasma in the electrostatic approximation.
Pressure effects are neglected for the ions whereas the electrons
are described by an isothermal equation of state. Contrary to the
quantum degenerate case,\cite{haas2} the present model is more
suitable to investigate the classical limit $\hbar\rightarrow0$.We
do not include quantum statistical effects in the present
investigation, and therefore, only quantum diffraction effects,
responsible e.g. for tunnelling, are taken into account.

The paper is organized as follows. In Section II we write the
quantum hydrodynamic model for a two-species plasma and derive the
Langmuir mode for quantum plasmas. In Section III, we obtain the
quantum Zakharov system through a procedure similar to the
classical one where a two-time scale formalism is used. In Section
IV we study the influence of quantum effects in two relevant
parametric instabilities: the decay instability and the four-wave
instability. Section VI is devoted to a preliminary discussion of
the nonlinear aspects of the problem and some of the remaining
open questions. Section V is reserved to the conclusions.

\section{Quantum Langmuir waves}
Before considering the nonlinear coupling between ion-acoustic
and Langmuir waves, we examine the linear stability analysis of the
Langmuir waves in the quantum regime.\cite{silin} For this
purpose we consider a one-dimensional quantum system, composed
of electrons and singly charged ions. The quantum hydrodynamic
equations in this case become\cite{haas2}
\begin{eqnarray}
\label{eq1}
\frac{\partial n_e}{\partial t} &+&
\frac{\partial(n_{e}u_{e})}{\partial x} = 0 \,, \\
\label{eqq1}
\frac{\partial n_i}{\partial t} &+&
\frac{\partial(n_{i}u_{i})}{\partial x} = 0 \,, \\
\label{eq2}
\frac{\partial u_e}{\partial t} &+&
u_{e}\frac{\partial u_e}{\partial x} = -
\frac{e}{m_e}E -\frac{1}{m_{e}n_{e}}\frac{\partial
P_{e}}{\partial x} +
\frac{\hbar^2}{2m_{e}^{2}}\frac{\partial}{\partial x}
\left(\frac{\partial^{2}\sqrt{n_e}/\partial
x^{2}}{\sqrt{n_e}}\right) \,, \\
\label{eqq2}
\frac{\partial u_i}{\partial t} &+&
u_{i}\frac{\partial u_i}{\partial x} =
\frac{e}{m_i}E \,, \\
\label{eq3} \frac{\partial E}{\partial x}
&=&\frac{e}{\varepsilon_0}(n_i - n_e) \,,
\end{eqnarray}
where $E$ is the electric field, $P_e$ is the electron pressure,
and $n_e$, $n_i$, $u_e$, $u_i$, $m_e$ and $m_i$ represent the
density, fluid velocity and mass of electrons ($e$) and ions
($i$), respectively. In addition, $\varepsilon_0$ and $\hbar$ are
the vacuum dielectric and the scaled Planck's constants. Since we
are interested in high frequency waves, the ion density $n_i$ can
be assumed constant, at this stage. The pressure $P_e$ is obtained
from an equation of state for the electrons, which basically
depends on the thermodynamic properties of the system. In the
present investigation, we consider the isothermal equation of
state $P_e=\kappa_B n_eT_e$, where $T_e$ is the electrons'
temperature and $\kappa_B$ is the Boltzmann's constant. In view of
their large mass, ions are treated classically. Also, in a first
approximation, we consider cold,  zero temperature ions. The Bohm
potential term proportional to $\hbar^2$ in (\ref{eq2}) is
responsible for negative differential resistance in semiconductor
devices\cite{10} and is associated to tunnelling.

Linearization of the electron equations (\ref{eq1}),
(\ref{eq2}) and (\ref{eq3}) around the homogeneous equilibrium
$n_e= n_i = n_0$, $u_e = 0$ and $E=0$ produces the following dispersion
relation:
\begin{equation}
\label{disp}
\omega^2=\omega_e^2+v_e^2k^2+{\hbar^2\over 4m_e^2}k^4\,.
\end{equation}
In Eq. (\ref{disp}), $\omega$ is the wave frequency, $k$ is the
wavenumber, $\omega_e=(n_0e^2/m_e\varepsilon_0)^{1/2}$ is the
electron plasma frequency and $v_e=(\kappa_BT_e/m_e)^{1/2}$ is the
electron thermal velocity. Notice that both classical and quantum
modes can be obtained from Eq. (\ref{disp}). In fact, the
classical limit $\hbar\rightarrow 0$ gives the classical Langmuir
wave dispersion relation.\cite{nichol} According to Eq.
(\ref{disp}), the frequency $\omega$ is always real, and
instability (or damping) of this wave cannot be observed. The main
purpose of this paper is to obtain a model describing the exchange
of energy between the quantum Langmuir modes shown above and the
recently found quantum ion-acoustic plasma modes.\cite{haas2}

\section{Quantum Zakharov equations}

In order to obtain the set of equations describing the
nonlinear interaction between Langmuir waves and ion-acoustic
waves, in the quantum regime, we follow the
derivation originally made by Zakharov.\cite{zak} A general
discussion of the validity of the Zakharov equations
can be found in the review paper by Thornhill and
ter Haar.\cite{thorn}

We first separate all fluid variables into high frequency
(subscript $h$) and low
frequency (subscript $l$) components,
\begin{eqnarray}
\label{eq4}
&&n_e(x,t)=n_0+n_{l}(x,t)+n_{h}(x,t)\,,\\
\label{i1}
&&n_i(x,t)=n_0+n_{l}(x,t)\,,\\
&&u_e(x,t)=u_{l}(x,t)+u_{h}(x,t)\,,\\
\label{i2}
&&u_i(x,t)=u_{l}(x,t)\,,\\
\label{eq8}
&&E(x,t)=E_l(x,t)+E_h(x,t)\,.
\end{eqnarray}
Notice that the high frequency portions of the ion quantities
[Eqs. (\ref{i1}) and (\ref{i2})]  were ignored due to the large
ion mass. Also, from the very beginning we assume that departures
from the quasi-neutral regime ($n_{i}\approx n_{e}$ and
$u_{i}\approx u_{e}$) are provided only by the high frequency
components of the electrons motion. The high frequency term of the
electric field can also be written as
\begin{equation}
\label{eh}
E_h(x,t)={1\over 2}\tilde{E}(x,t)e^{-i\omega_et} +
\mbox{c.c.}\,,
\end{equation}
where $\tilde{E}(x,t)$ is the slowly varying envelope of the high
frequency term and c.c. refer to complex conjugate. Using the high
frequency components of Eqs. (\ref{eq1}-\ref{eq3}), we obtain, by
the same procedure used in the classical case,\cite{thorn}
\begin{equation}
\label{zha1}
i{\partial \tilde{E}\over\partial t}+{1\over2}{v_e^2\over\omega
_e}
{\partial ^2\tilde{E}\over\partial x^2}-{\hbar
^2\over8m_e^2\omega _e}
{\partial ^4\tilde{E}\over\partial x^4}={\omega
_e\over2}{n_{l}\over n_0}\tilde{E}\,,
\end{equation}
where the term $|\partial^2_t\tilde{E}|\ll
|\omega_e\partial_t\tilde{E}|$ has
been neglected. Equation (\ref{zha1}) describes the evolution
of the slowly varying amplitude $\tilde{E}$, as defined in (\ref{eh}).

We next proceed with the derivation of the equation for the low
frequency part, $n_l$, of the departure from the equilibrium
density $n_0$. After averaging over the fast time scale, we get a
set of equations describing the low frequency part of the electron
continuity equation, electron force equation and ion force
equation,
\begin{eqnarray}
\label{eq12}
&&{\partial n_{l}\over\partial t}+n_0{\partial
u_{l}\over\partial x}=0\,,\\
\label{eq13}
&&{\partial u_{l}\over\partial t}+{e\over m_e}E_l+{\kappa _B
T_e\over n_0 m_e}
{\partial n_{l}\over\partial x}
-{\hbar^2\over4m_e^2n_0}{\partial ^3n_{l}\over\partial x^3}
+{e^2\over4m_e^2\omega_e^2}{\partial |\tilde{E}|^2\over\partial
x}=0\,,\\
\label{eq14}
&&{\partial u_{l}\over\partial t}-{e\over m_i}E_l=0\,.
\end{eqnarray}
Convective terms were disregarded in view of a weak Langmuir
turbulence assumption, as detailed by Thornhill and ter Haar.\cite{thorn}
Eliminating $u_{l}$ and $E_l$ from Eqs. (\ref{eq12}-\ref{eq14}) and assuming
$m_e/m_i\ll 1$, we obtain
\begin{equation}
\label{zha2}
{\partial ^2n_{l}\over\partial t^2}-c_s^2{\partial ^2
n_{l}\over\partial x^2}
+{\hbar^2\over4m_im_e}{\partial ^4n_{l}\over\partial x^4}
={\varepsilon_0\over4m_i}{\partial ^2|\tilde{E}|^2\over\partial
x^2}\,,
\end{equation}
where $c_s=(\kappa_BT_e/m_i)^{1/2}$ is the ion-acoustic
velocity. We call
Eqs. (\ref{zha1}) and (\ref{zha2}) the {\em quantum Zakharov
equations}.

For the following analysis, it is most convenient to normalize
Eqs.
(\ref{zha1}) and (\ref{zha2}). Normalized quantities are
expressed as
\begin{eqnarray}
\label{eq15}
\bar{x} &=& 2\,\sqrt{{m_e\over m_i}}\,{x\over\lambda_e} \,,
\quad
\bar{t} =2{m_e\over m_i}\,\omega_{e}\,t \,, \\
\label{eq16}
\bar{n} &=& {1\over4}{m_i\over m_e}{n_{l}\over n_0} \,, \quad
\bar{E}=\sqrt{{\varepsilon_0\,m_i\over16\,m_e\,n_0\,\kappa_B\,T_e}}\tilde{E}
\,,
\end{eqnarray}
where $\lambda_e$ is the electron Debye length.
In addition to (\ref{eq15}-\ref{eq16}), we introduce the
dimensionless
quantum parameter
\begin{equation}
\label{eq17}
H={\hbar\, \omega_{i}\over\kappa_B\,T_e}\,,
\end{equation}
where $\omega_i = (n_{0}e^{2}/m_{i}\varepsilon_{0})^{1/2}$ is
the
ion plasma frequency. The resulting system reads (we dropped
bars
for the sake of simplicity)
\begin{eqnarray}
\label{zn1}
&&i{\partial E\over\partial t}+{\partial ^2E\over\partial
x^2}-H^2
{\partial ^4E\over\partial x^4}=n\,E\,,\\
\label{zn2}
&&{\partial ^2n\over\partial t^2}-{\partial ^2 n\over\partial
x^2}
+H^2{\partial ^4n\over\partial x^4}={\partial
^2|E|^2\over\partial x^2}\,.
\end{eqnarray}
The quantum parameter $H$ given in (\ref{eq17}) expresses the
ratio between the ion plasmon energy and the electron thermal
energy. This is to be compared with the dimensionless parameter
characterizing quantum effects in the two-stream quantum
instability,\cite{hmf} given by the ratio between electron
plasmon and thermal energies. Here, the presence of ion-acoustic
modes forces the appearance of ionic (inertia) parameters. Notice
that for dense plasmas,\cite{jung} with particle density about
$10^{25}-10^{32}\, \mbox{m}^{-3}$ and temperature about
$10^5-10^7\, \mbox{K}$, the parameter $H$ is not irrelevant at
all. For a completely ionized hydrogen plasma in these ranges of
densities and temperatures, $H$ typically goes from negligible
values of order $10^{-5}$ up to values of order unity. The
presence of large values of $H$ points to the possible
experimental manifestation of quantum effects in the coupling
between Langmuir and ion-acoustic modes in dense plasmas,
particularly in astrophysical plasmas.

In the next section, the model (\ref{zn1}-\ref{zn2}) is used
to investigate two parametric instabilities extensively
studied in classical plasmas: the decay instability and the
four-wave interaction.

\section{Parametric instabilities}

\subsection{Decay instability}
Following strictly the treatment for the classical decay
instability,\cite{thorn} consider the proposal
\begin{eqnarray}
\label{lang} E &=& E_{0} e^{i(k_{0}x-\omega_{0}t)} + E_{1}(t)
e^{i(k_{1}x-\omega_{1}t)} \,,\\
\label{ionac} n &=& n_{1}(t) \cos(Kx - \Omega t) \,,
\end{eqnarray}
now for the quantum Zakharov equations (\ref{zn1}-\ref{zn2}),
where $E_{1}(t)$ and $n_{1}(t)$ are first-order quantities, and
\begin{eqnarray}
\label{d1}
\omega_0 &=& k_{0}^2 + H^{2}k_{0}^4 \,, \\
\label{d2}
\omega_1 &=& k_{1}^2 + H^{2}k_{1}^4 \,, \\
\label{d3} \Omega^2 &=& K^2 + H^2 K^4 \,.
\end{eqnarray}
Notice that Eq. (\ref{d3}) is identical to the quantum dispersion
relation obtained by Haas {\em et al.}\cite{haas2} (Section V) except for the value
of $H$ which, due to the use of a quantum equation of state, has a
different definition there.

Also, there are the usual matching conditions
\begin{equation}
\label{coup} k_0 = k_1 + K \,, \quad \omega_0 = \omega_1 +
\Omega
\,,
\end{equation}
corresponding to momentum and energy conservation respectively.
These conditions describe the decay of one quantum Langmuir wave,
with dispersion relation (\ref{d1}), into other quantum Langmuir
wave, with dispersion relation (\ref{d2}), and a quantum
ion-acoustic wave, with dispersion relation (\ref{d3}).

Linearizing the quantum Zakharov equations, we obtain
\begin{eqnarray}
\label{z1}
&&i\dot{E_1} e_1 = \frac{E_{0}n_1}{2}(e_{+} + e_{-})e_0 \,, \\
\label{z2} &&\left(\frac{\ddot{n_1}}{2}\right. -
i\Omega\dot{n_1}
+ \left. K^{2}E_{0}E_{1}^{*}\right)e_{+} +
\left(\frac{\ddot{n_1}}{2} + i\Omega\dot{n_1} +
K^{2}E_{0}^{*}E_{1}\right)e_{-} = 0 \,,
\end{eqnarray}
where use has been made of the notation
\begin{equation}
e_{0,1} = \exp[i(k_{0,1}x - \omega_{0,1}t)] \,, \quad e_{\pm} =
\exp[\pm i(Kx - \Omega t)] \,.
\end{equation}

The resonant part [satisfying (\ref{coup})] of (\ref{z1}) gives
\begin{equation}
n_1 = \frac{2i}{E_0}\dot{E_1} \,,
\end{equation}
while (\ref{z2}) gives
\begin{equation}
\frac{\ddot n_1}{2} + i\Omega\dot{n_1} + K^{2}E_{0}^{*}E_{1} =
0
\,.
\end{equation}

For $\dot{E_1} = i\omega E_1$, elimination of $n_1$ leaves us
with
\begin{equation}
\label{de} \omega^3 + 2\Omega\omega^2 + K^{2}|E_0|^2 = 0 \,,
\end{equation}
which is formally identical to the dispersion relation for the
classical decay instability. Hence, all conclusions valid for
the classical case can be immediately extended to the quantum case.
In particular, for $\Omega \gg |\omega|$, so that the cubic term
can be neglected in (\ref{de}), and for $\omega = i\gamma$, we
obtain the growth rate
\begin{equation}
\label{gam}
\gamma = \frac{K|E_0|}{\sqrt{2\Omega}} \,.
\end{equation}
In all situations, the discriminant\cite{rhyzik} of the cubic
equation (\ref{de}) is positive and there are one real and two
complex conjugate solutions for this equation, one of which is
necessarily unstable.

To conclude, there is a formal similarity between the classical
and quantum decay instabilities. The only differences remain in
the dispersion relations (\ref{d1}-\ref{d3}), for the quantum
Langmuir and ion-acoustic modes.
%
%
The quantum dispersion relations, however, produces  a saturation
effect not present in the classical case (see Fig. 1). Combining
(\ref{d3}) and (\ref{gam}), we obtain
\begin{equation}
\gamma = \frac{\sqrt{K}|E_0|}{\sqrt{2}(1 + H^{2}K^{2})^{1/4}} \,,
\end{equation}
for which a maximum value $\gamma_{max} = |E_{0}|/\sqrt{2H}$ is
obtained for $K \rightarrow \infty$. This is to be compared with
the classical case ($H = 0$) where $\gamma$ grows with no bound as
$K$ increases. Even if the quantum effects do not imply
stabilization, they limit the instability to a fixed maximum
growth rate.

For dense astrophysical hydrogen plasmas,\cite{jung} where $n_0
\sim 10^{32} m^{-3}$ and $T_e \sim 10^5 K$, we obtain $H \sim 1$
and $\gamma_{max} = |E_{0}|/\sqrt{2}$. For laser hydrogen plasmas,
however, to the best of our knowledge, such high densities are not
yet attainable. For today's typical values\cite{bonitz} of $n_0
\sim 10^{28} m^{-3}$ and $T \sim 10^{5} K$, we obtain a modest
value of $H \sim 10^{-2}$. The smallness of the quantum effects
for {\it modulational} instabilities in laser plasmas follows from
the fact that, in this respect, the relevant parameter $H$ is
defined as the ratio of the ion plasmon energy to the electron
thermal energy. If the pertinent quantum parameter were the ratio
between the {\it electron} plasmon energy and the electron thermal
energy, as in the quantum two-stream instability,\cite{hmf} the
lower mass of the electrons would increase significantly $H$. For
laser plasmas with the same typical values as before, we would
have $H \sim 0.4$. Of course, we are not saying that quantum
effects are irrelevant for laser plasmas: they can show up when
the Landau length becomes comparable to the de Broglie wavelength,
in which case $\hbar\omega_c \sim \kappa_{B}T_e$, where $\omega_c$
is the cyclotron frequency associated with the laser field, or
when the electrons are degenerated.\cite{bonitz}
%
%

\subsection{Four-wave instability}
\label{four}
The general instability involving the
interaction between one single finite-amplitude Langmuir wave,
two other Langmuir waves and one ion-acoustic wave
can be obtained by choosing\cite{nichol}
\begin{eqnarray}
\label{eq18}
E(x,t)&=&E_0\exp(-i\omega_0t+ik_0x)+E_{+}\exp[-i(\omega_0+\omega)t
+i(k_0+k)x]\nonumber\\
&+&E_{-}\exp[-i(\omega_0-\omega^*)t+i(k_0-k)x]\,,\\
\label{eq19}
n(x,t)&=&\tilde{n}\exp(-i\omega t+ikx)+\mbox{c.c}\,,
\end{eqnarray}
where the amplitudes $E_+$, $E_-$ and $\tilde{n}$ are all first
order quantities. We choose the equilibrium solution
$E(x,t)=E_0\exp(-i\omega_0t+ik_0x),$ $n(x,t)=0$ to satisfy Eqs.
(\ref{zn1}-\ref{zn2}) with $E_0$ real. This implies the relation
\begin{equation}
\label{disp1}
\omega_0=k_0^2 + H^2 k_0^4\,.
\end{equation}
The last equation can be found also from the dispersion relation
for quantum Langmuir modes [equation (\ref{disp})] taking
$\omega_e = 0$ and the recalling of variables. In fact, $\omega_e$
is absorbed in the electric field through (\ref{eh}). In
conclusion, (\ref{disp1}) shows a quantum Langmuir mode.

The forms (\ref{eq18}-\ref{eq19}) when inserted in Eq.
(\ref{zn1})
yield
\begin{eqnarray}
\label{eq20}
&&(\omega_0+\omega)E_+-(k_0+k)^2E_+=\tilde{n}E_0+H^2(k_0+k)^4E_+\,,\\
\label{eq21}
&&(\omega_0-\omega^*)E_--(k_0-k)^2E_-=\tilde{n}^*E_0+H^2(k_0-k)^4E_-\,.
\end{eqnarray}
When combined, Eqs. (\ref{zn2}), (\ref{disp1}-\ref{eq20}) and
the complex conjugate
of (\ref{eq21}) give the following dispersion relation:
\begin{equation}
\label{disp2}
 D_sD_1D_2=k^2E_0^2(D_1 + D_2)\,,
\end{equation}
in which
\begin{eqnarray}
\label{Ds}
&&D_s = \omega^2-k^2-H^2k^4\,,\\
\label{D1}
&&D_1 =
\omega-k^2-2k_0k-H^2(k^4+4k_0k^3+6k_0^2k^2+4k_0^3k)\,,\\
\label{D2}
&&D_2=-\omega-k^2+2k_0k-H^2(k^4-4k_0k^3+6k_0^2k^2-4k_0^3k)\,.
\end{eqnarray}
Notice that the limit $H\rightarrow 0$ recovers the classical
dispersion relation for the four-wave interaction.\cite{nichol}

The dispersion relation (\ref{disp2}) is a fourth order
polynomial in
$\omega$ that, in general, can only be analyzed numerically.
However, the simplest case $\omega_0=k_0=0$ can be investigated
analytically. For a
purely growing instability $\omega=i\gamma$ the dispersion
relation (\ref{disp2}) becomes
\begin{equation}
\label{disp3}
[\gamma^2+k^2+H^2k^4][\gamma^2+(k^2+H^2k^4)^2]=2k^2E_0^2(k^2+H^2k^4)\,.
\end{equation}
Solving for $\gamma^2$, there follows two roots,
\begin{eqnarray}
\label{xxx}
\gamma^2 &=& - \frac{1}{2}(k^2 + H^2 k^4)(1 + k^2 + H^2 k^4) \\
&\pm& \frac{1}{2}(k^2 + H^2 k^4)^{1/2}[(k^2 + H^2 k^4)(1 - k^2
- H^2 k^4)^2
+ 8 k^2 E_{0}^2]^{1/2} \,, \nonumber
\end{eqnarray}
one of them necessarily negative (stable mode). The other root
is positive (unstable mode) provided
\begin{equation}
E_{0}^2 > \frac{k^2}{2}(1 + H^2 k^2)^2 \,.
\end{equation}
This recovers the classical instability condition for the
four-wave interaction\cite{nichol} when $H \rightarrow 0$.
However, there is a new, quantum effect of instability
suppression provided
\begin{equation}
H^2 \geq \frac{\sqrt{2}E_0 - k}{k^3} \,.
\end{equation}
This is in agreement with the overall stabilization that
quantum diffraction effects produce in high density
plasmas.\cite{hmf,suh} In fact, for sufficiently large $H$
there is no transfer of energy from the original quantum Langmuir
mode to the two new quantum Langmuir modes and to the
quantum ion-acoustic mode.

Let us consider in more detail the potentially unstable mode
described by the positive root in (\ref{xxx}). In Fig. 2, we show
$\gamma^2$ versus $k^2$ for $H =0$, $H = 0.5$ and $H = 0.9$, with
$E_0 = 0.5$. The instability region ($\gamma^2>0$) in $k$-space
becomes narrower for bigger $H$. Also, the maximum $\gamma^2$
becomes smaller the larger the quantum effects. This latter result
is analytically supported by an expansion of the positive root of
Eq. (\ref{xxx}). The wavenumber for maximum growth rate, $k_{max}$
can be calculated, in a $O(k^5)$ approximation, by expanding Eq.
(\ref{xxx}) up to fifth order in $k$. This perturbation analysis
is interesting since $d\gamma/dk = 0$ is not soluble in closed
form if we use the exact expression for (\ref{xxx}). The result of
the expansion procedure is shown in Fig. 3, where the wavenumber
$k_{max}$ for maximum growth rate when $E_0 = 0.5$ is shown as a
function of $H$. (Notice the extended domain of the function,
beyond the reasonable limit of $H \cong 1$.) Using this $k_{max}$
we obtain a somewhat complicated expression which can be used to
calculate the associated growth rate $\gamma$. Using a computer
algebra program, we can easily obtain $\gamma_{max} = E_{0}^{2}(1
- E_{0}^2 - 0.87 H^2)$, a result valid up to $O(E_{0}^4)$. This is
an approximate equation showing that quantum effects produce
stabilization. The approximations adopted are justified in view of
our assumptions of long wavelengths and weak turbulence (small
electric field amplitudes).
%
%
For dense  astrophysical plasmas with $H \sim 1$, as in the decay
instability case, we would get $\gamma_{max} = E_{0}^{2}(0.13 -
E_{0}^{2})$, a significant difference in comparison with the
classical case where $\gamma_{max} = E_{0}^{2}(1 - E_{0}^{2})$.
%
%

In order to further assess the role of quantum effects in the
four-wave interaction process, we performed a numerical study of
(\ref{disp2}) for general $k_0 \neq 0$. Figure 4 displays the real
(solid lines) and imaginary (dashed lines) parts of $\omega$ as a
function of $k$. Both uncoupled (i.e., $E_0\approx 0$) and coupled
cases are considered, for three different values of $H$. Due to
the symmetry ($k,\omega) \leftrightarrow (-k,-\omega)\Rightarrow
D_1\leftrightarrow D_2$, of the dispersion relation (\ref{disp2}),
we consider only positive values of the wavenumber, around the
overlay region of the branches $D_s$ and $D_2$, where instability
occurs. In the uncoupled case, $k=2k_0$ is a root of $D_2$ when
$\omega = 0$, for both classical and quantum cases. Also, the
plots of $D_s$ and $D_2$ branches touch each other at isolated
points while, when $E_0 \neq 0$, overlay occurs for a whole finite
interval of $k$, signalizing wave instability. The first column of
plots shows that, for a fixed $k_0$, both uncoupled curves raise
with $H$, implying reduction of the interval in $k$ where
instability settles down. This can be checked against the
corresponding figures in the second and third columns, where a
contraction of the unstable interval is clearly seen.

Denote the unstable interval in $k$ by $I_k = (k_a, k_b)$. For
higher pump energy $E_0$, the third column of Fig. 4 shows an
overall contraction of $I_k$. This results from the gradual shift
of $k_a$ to the right and  $k_b$ to the left, due to the quantum
effects. For the relevant range of values $0 \leq H \leq 1$, less
severe attenuations occur for the maximum growth rate, compared to
those found for the unstable interval in $k$. Thus, the numerical
results show that the quantum effect inhibits the spreading of
energy among different modes. In fact, assume that for a specific
$k$, $N_I = (k_b-k_a)/k$ represents a first estimation for the
number of active modes at the beginning of the process. Then, the
contraction of $I_k$ implies that the Langmuir fluctuations in
quantum plasmas might represent more coherent configurations,
i.e., having less effective modes when compared to the
corresponding classical situation, an issue to be checked by a
direct numerical simulation and, possibly, by an experiment.

\section{Nonlinear analysis and open questions}
An important regime of the classical Zakharov equations
concerns its static limit. In this case, the classical Zakharov system
do possess soliton solutions described by a nonlinear Sch\"odinger
equation.\cite{thorn} The procedure for the static limit of
the quantum Zakharov equations considers the approximation
$\partial^{2}n/\partial t^2 \approx 0$ in (\ref{zn2}). This
gives immediately
\begin{equation}
\label{sta1}
n = - |E|^2 + H^{2}\frac{\partial^{2}n}{\partial x^2} \,.
\end{equation}
Equation (\ref{sta1}), inserted in Eq. (\ref{zn1}), yields
\begin{equation}
\label{sta2}
i\frac{\partial E}{\partial t} + \frac{\partial^{2}E}{\partial
x^2} + |E|^{2}E =
H^{2}\left(\frac{\partial^{4}E}{\partial x^4} +
E\frac{\partial^{2}n}{\partial x^2}\right) \,.
\end{equation}
In the classical limit $H\rightarrow 0$, the right-hand side of
Eq. (\ref{sta2}) vanishes and we recover the  nonlinear
Schr\"odinger equation with its soliton solutions. In the
quantum case, however, equations (\ref{sta1}-\ref{sta2}) form a
coupled, nonlinear system. We have not been able to find localized,
analytical solutions for this system. In fact, the usual
reduction procedure of searching for solutions in the form
\begin{equation}
E = F(x-Mt)\exp(i[k(x-ut)+\delta]) \,, \quad  n = G(x-Mt) \,,
\end{equation}
for real $F$, $G$, $k$, $M$, $u$ and $\delta$ produces a
complicated fourth-order system of coupled, nonlinear
equations. The existence of soliton solutions for this system remains an
open question. It seems that a numerical analysis could help in this
respect but we believe that this issue should be more
appropriately treated in a future work.

Another avenue in nonlinear studies of the quantum Zakharov
equation concerns its simultaneous semiclassical and static
limit. Substituting (\ref{sta1}) into (\ref{sta2}) and retaining only
terms up to $O(H^2)$ produces the decoupled equation
\begin{equation}
\label{nlsq}
i\frac{\partial E}{\partial t} + \frac{\partial^{2}E}{\partial
x^2} + |E|^{2}E =
H^{2}\left(\frac{\partial^{4}E}{\partial x^4} -
E\frac{\partial^{2}|E|^{2}}{\partial x^2}\right) \,.
\end{equation}
Equation (\ref{nlsq}) can be used to study perturbations of the classical
NLS soliton solutions. The terms proportional to $H^2$, in Eq. (\ref{nlsq}),
will probably modify the dispersion-nonlinearity equilibrium, which
is the ultimate responsible for the soliton existence.

More formal aspects of the Zakharov equations have to do with its
variational formulation and the associated Noether currents.\cite{gib}
In particular the quantum Zakharov equations preserve
the number of plasmons $\int |E|^{2} dx$ of the high frequency
electric field, as a consequence of the associated conservation
law
\begin{equation}
\label{contin}
 \frac{\partial\rho}{\partial t} + \frac{\partial
J}{\partial x} = 0 \,,
\end{equation}
where $E(x,t) = A(x,t) \exp(i\theta(x,t))$, with $A = A(x,t)$
and $\theta = \theta(x,t)$ real amplitude and phase functions, and
\begin{eqnarray}
\rho &=& A^2 \,, \\
J &=& 2\,A^2 \frac{\partial\theta}{\partial x} - 2\,H^{2}\left[A^2
\frac{\partial^{3}\theta}{\partial x^3} + 2\,A \frac{\partial
A}{\partial x}
 \frac{\partial^{2}\theta}{\partial x^2}\right. \\
 &-& 2\left.
A^2\left(\frac{\partial\theta}{\partial x}\right)^3 -
2\left(\frac{\partial A}{\partial
x}\right)^{2}\frac{\partial\theta}{\partial x} + 4\, A
\frac{\partial^{2}A}{\partial x^2} \frac{\partial\theta}{\partial
x}\right] \,. \nonumber
\end{eqnarray}
Notice the extra contribution proportional to $H^2$ to the
plasmons current.The conservation law (\ref{contin}) comes from
the imaginary part of (\ref{zn1}) and hence contains no
contribution from $n$. A proper formulation of the remaining
conservation laws (momentum and energy) of the system is an open
question to be tackled, preferably in accordance with symmetry
principles of an associated action functional. Other important
issues concern the search for coherent solutions of the quantum
Zakharov equations, namely quantum solitons and quantum cavitons.

Still another issue related to the nonlinear analysis of Eqs.
(\ref{zn1}-\ref{zn2}) concerns thermalization and recurrence. For
periodic boundary conditions, the classical NLS does not exhibit
thermalization and, therefore, is generically
recurrent.\cite{Tyagaraja79,Tyagaraja81} The classical procedure to address
such questions is based on estimations for the number of active
modes $N_A$, from the Rayleigh quotient. An upper bound estimation
for this number is provided by two invariants: the number of
plasmons and a momentum--like invariant which, in our case, is not
yet known. For classical regimes, numerical simulations show that
the conclusions can, in general, be extended to the non-integrable
Zakharov system, when considered as a perturbation of the NLS
regime.\cite{Oliveira97} In fact, it has been shown that, at
least for some period of time, the constancy of the momentum-like
quantity is approximately satisfied. Moreover, numerical
simulations show that the elementary estimation presented in the
last section, i.e., $N_A\approx N_I$, can yield quite good results
when applied to the full Zakharov equations. Under this viewpoint,
the contraction of the $k$-unstable interval due to $H\neq 0$,
verified in subsection (\ref{four}), suggests that the
distribution of energy is less intense in quantum plasmas when
compared with the classical case. Therefore, quantum effect would
favor recurrence in Langmuir modulational regimes.

To finalize, we can derive some exact solutions for the quantum
Zakharov equations (\ref{zn1}-\ref{zn2}) if we consider pure
ion-sound waves obtained by taking $E = 0$. With zero electric
field, the density perturbation satisfies the undriven equation
\begin{equation}
\label{pure}
\frac{\partial^{2}n}{\partial t^2} - \frac{\partial^{2}n}{\partial x^2}
+ H^{2}\,\frac{\partial^{4}n}{\partial x^4} = 0 \,.
\end{equation}
This linear fourth-order evolution equation was investigated using
the method of Lie symmetries\cite{olver} and we found time and
space translation symmetries, as well as a scale symmetry
resulting from the linearity. The $H^2$ term breaks down the
Lorentz invariance endowed by the classical model for pure
ion-sound waves, so that arbitrary waves travelling at the
ion-sound velocity can not be constructed. Nevertheless, exact
solutions for Eq. (\ref{pure}) can be found supposing  $n =
\bar{n}(x- ct)$, for constant $c$ and for $\bar{n}$ a function to
be determined. For $c^2 > 1$, corresponding to supersonic flow,
and disregarding an integration constant associated to non-bound
solutions, we get periodic solutions of the form
\begin{equation}
n = a + b \cos\left(\frac{\sqrt{c^2 - 1}}{H}\,(x - ct) + \delta\right) \,,
\end{equation}
where $a$, $b$ and $\delta$ are numerical constants. This
similarity solution is an arbitrary amplitude solution. Notice
that quantum effects increase the spatial frequency of
oscillations in the reference frame of the travelling wave.

\section{Conclusion}
We obtained a general model to analyze the coupling between
Langmuir waves and ion-acoustic waves, in a quantum setting. The
model was shown to be appropriate to the four-wave interaction and
quantum effects have been shown to provide stabilization of a
classically unstable mode. In the case of the decay instability, a
formal similarity with the classical case is identified, except
for small differences in the dispersion relations, representing
quantum corrections. We also identified a dimensionless quantum
parameter given by the ratio of the ion plasmon and electron
thermal energies. As pointed out before, this quantum parameter
may not be small, at least for dense plasmas.

%
%
The consequences of our results on today´s laboratory or
technological plasmas are not yet fully assessed since, for
present conditions, $H \ll 1$ in these applications. However
quantum effects may imply important consequences in the behavior
of high density astrophysical plasmas, where $H \sim 1$ is easily
found. In this case, as we pointed out, quantum effects cause an
overall reduction in the wave-wave interaction level. Specifically
and in contrast to the classical case, the decay instability
growth rate is bounded for large wavenumbers. Growth rate
reduction also occurs for the four-wave interaction. Besides,
suppression is also verified in the length of the unstable
spectral range, implying spectral focusing, i.e., a restriction on
the range of possible unstable wave-numbers. This focusing effect
may extend to quite long periods of time, indicating that the
recurrence properties verified in the classical Zakharov equation
are enhanced by the quantum effects.
%
%

A number of open questions remains to be addressed. Of course, a
complete analysis of the linear dispersion relation of the quantum
Zakharov system have to be done. This may require a full
three-dimensional treatment, with the inclusion of electromagnetic
coupling between Langmuir and ion-acoustic modes. An additional
important point are the nonlinear effects, some of them briefly
discussed in Section V, which may deserve a more careful scrutiny.
To conclude, the huge amount of physical and mathematical aspects
already assessed in the classical Zakharov equations certainly
have quantum counterparts which ask for an equally careful
investigation.

\bigskip \noindent{\bf Acknowledgments}\\ This work was partially
supported by Conselho Nacional de Desenvolvimento
Cient\'{\i}fico e Tecnol\'ogico - CNPq. One of us (L.~G.~G.) gratefully
acknowledges Universidade do Vale do Rio dos Sinos - UNISINOS for
hospitality and support during the preparation of this work.

\newpage
\centerline{FIGURE CAPTIONS} \vskip 3cm

%
%
FIG. 1. Growth rate of the decay instability for  $E_0=0.5$ and $0
\le H \leq 1.5$, as indicated. Notice the quick saturation effect
for $H > 0$. \vskip 2cm
%
%

FIG. 2. $\gamma^2$ as a function of $k^2$ for the positive root in
the dispersion relation (\ref{xxx}) for the four--wave
instability. We have $E_0 = 0.5$, $H = 0$ (full line), $H = 0.5$
(dashed line) and $H = 0.9$ (dotted line).
 \vskip 2cm

FIG. 3. Wave-number $k_{max}$ for maximum growth rate of the
four-wave instability, as a function of $H$, calculated to
$O(k^5)$ and $E_0 =0.5$. \vskip 2cm

FIG. 4. Real (solid lines) and imaginary (dashed lines) components
of the frequency $\omega$ as a function of $k$ for uncoupled
(frames {\em a1}, {\em a2} and {\em a3}) and coupled cases (frames
{\em b1} to {\em c3}). From top to bottom, $H = 0$, $H = 0.5$ and
$H = 0.9$, respectively. From left to right, $E_0 = 0$, $E_0 =
0.5$ and $E_0 = 0.5$. For the first and second columns $k_0=0.5$;
for the third column, $k_0=0.75$. In the first frame, $D_i$ ($i=
s, 1, 2$) indicate the various branches of (\ref{disp2}). A
similar labelling applies to all the frames.
 \vskip 2cm

\end{document}